\documentclass[sigconf,anonymous=false]{acmart}
\renewcommand\footnotetextcopyrightpermission[1]{}
\AtBeginDocument{%
  }

\setcopyright{acmlicensed}
\copyrightyear{2025}
\acmYear{2025}
\acmDOI{XXXXXXX.XXXXXXX}
\acmConference[Conference]{ International Conference for High Performance Computing, Networking, Storage, and Analysis}{Nov 15-20, 2026}{Chicago, IL}

\acmISBN{978-1-4503-XXXX-X/18/06}
\usepackage{mdframed}
\usepackage{svg}

\usepackage{xspace}
\usepackage{paralist}

\usepackage{caption}
\usepackage{subcaption}

\usepackage{mdframed}
\usepackage[T1]{fontenc} % For proper font encoding
\usepackage{minted} % For syntax highlighting
\usepackage{xcolor} % For custom colors, if desired

\usepackage{tabularx}
\usepackage{threeparttable} % For table notes
\usepackage{multirow}
\usepackage{tabularray}
\usepackage[table]{xcolor}% http://ctan.org/pkg/xcolor
\usepackage{tabularx}
\usepackage[dvipsnames]{xcolor}
\usepackage{makecell}
\usepackage[acronym,nomain,nonumberlist]{glossaries}
\usepackage{booktabs} % For professional-looking rules
\usepackage{siunitx}  % For number formatting and alignment

\usepackage[font=small,labelfont=bf,skip=2pt]{caption}
\makeglossaries

\newacronym{ai}{AI}{artificial intelligence}
\newacronym{ci}{CI}{continuous integration}
\newacronym{bsp}{BSP}{bulk synchronous parallel}
\newacronym{mpi}{MPI}{Message Passing Interface}
\newacronym{hpc}{HPC}{high performance computing}
\newacronym{aws}{AWS}{Amazon Web Services}
\newacronym{gke}{GKE}{Google Kubernetes Engine}
\newacronym{aks}{AKS}{Azure Kubernetes Service}
\newacronym{eks}{EKS}{Elastic Kubernetes Service}
\newacronym{aiml}{AI/ML}{Artificial Intelligence and Machine Learning}
\newacronym{ml}{ML}{machine learning}
\newacronym{rdma}{RDMA}{Remote Direct Memory Access}
\newacronym{os}{OS}{operating systems}
\newacronym{vm}{VM}{virtual machine}
\newacronym{llm}{LLM}{Large Language Models}
\newacronym{fom}{FOM}{figure of merit}
\newacronym{efa}{EFA}{Elastic Fabric Adapter}
\newacronym{ec2}{EC2}{Elastic Compute Cloud}
\newacronym{ucx}{UCX}{Unified Communication X}
\newacronym{cni}{CNI}{container networking interface}
\newacronym{ebpf}{eBPF}{extended Berkeley Packet Filter}
\newacronym{gvnic}{gVNIC}{Google Virtual NIC}
\newacronym{uri}{URI}{unique resource identifier}
\newacronym{api}{API}{application programming interface}
\newacronym{dag}{DAG}{directed acylcic graph}
\newacronym{crd}{CRD}{custom resource definition}
\newacronym{mcp}{MCP}{Model Context Protocol}
\newacronym{hci}{HCI}{Human-Computer Interaction}
 
\newacronym{pid}{PID}{Process IDentifier}

\begin{document}

% Idea from Tapasya:
% survey of performance and usability of HPC applications in cloud
\title{Descriptive Execution of HPC Applications and Workflows}
% \title{Cross Cloud Performance Study\protect\\ Informing Possible Futures for HPC\protect\\}

% \subtitle{\large(1) Paper Type: Regular}


\author{Vanessa Sochat}
\authornote{Corresponding Author}
\email{sochat1@llnl.gov}
\orcid{0000-0002-4387-3819}
\affiliation{%
  \institution{Lawrence Livermore National Laboratory}
  \city{Livermore}
  \state{California}
  \country{USA}
}

\author{Daniel Milroy}
\email{milroy1@llnl.gov}
\orcid{0000-0001-6500-3227}
\affiliation{%
  \institution{Lawrence Livermore National Laboratory}
  \city{Livermore}
  \state{California}
  \country{USA}
}

\renewcommand{\shortauthors}{Sochat and Milroy.}

\begin{abstract}
The means to execute and orchestrate software components has changed from human-written code to descriptive prose. In high performance computing, this transition is represented in application orchestration, workload management, and system monitoring and debugging, to name a few. The underlying means to enable descriptive definition of tasks is the use of Large Language Model with associated tool functions and resources. A combination of a model with access to such resources, modeled in software, encompasses an autonomous framework. As fully automated and agentic frameworks are developed for science, it is important to assess reliability and strategies scoped to specific tasks. In this work, we assess the extent to which an agentic framework can  optimize and run an HPC scaling study with a low latency network in Amazon Web Services, accurately transform HPC job specifications between workload managers, and design and run an entire biosciences workflow. We find that the framework completes all three tasks while surfacing task-specific failure modes. In the scaling study, agents deploy and optimize applications but monitor running jobs inefficiently, preferring conservative fixed waits over event subscriptions. In job translation, they convert specifications between Slurm and Flux with high accuracy, with processor-affinity flags the most common error. In the bioscience workflow, the agent reproduces an expert-written variant-calling pipeline almost exactly---agreeing with the reference call set in 18 of 19 completed runs---and reaches this result through many distinct yet functionally equivalent workflow implementations. This information is invaluable moving forward to developing multi-cluster setups with scheduling and transformation handled by agents.
\end{abstract}

\maketitle
\section{Introduction}

The need for representation of tasks into inputs that can processed by a machine can be traced back as far as the 1890 U.S. Census, when punch cards were used to automate voting \cite{punchcards}. The method expanded in the 1900s to handle more tasks, from serving as library cards to proving mathematical theorems, and were foundational to modern computing today. The study of how people communicate with computers is \gls{hci} \cite{Marsh1990-vw} and is based on the idea of mapping human intent into tasks that can be executed by a computer. Software is an attempt to materialize human understanding into operations that a computer can perform. Data captures states of phenomena that, for scientific contexts, we typically want to use to model the real world. A data format is an attempt to standardize the phenomena into structures that can  best be processed by precise instructions. A modern workflow is capturing and executing a set of tasks with states, and dependencies. Even in early computing, insert of a punch card or execution of code was a rudimentary conversation between man and machine. A successful interaction leads to a meaningful computational result that is aligned with the initial intent.

The different strategies that computational scientists have adopted to use software and data to accomplish scoped tasks can be described as progressive stages. Andrej Kaparthy defines Software 1.0, 2.0, and 3.0 \cite{Pajo2025-ms} as a transition through stages of traditional coding, neural network methods, and using \gls{llm}, respectively. We might consider units of execution changing from single programs to workflows and now conversational \gls{ai}. Tasks are delegated to agents instead of programs. The 1960s through 1990s were defined by direct management of hardware paired with command line or software interfaces \cite{HEMPEL199951, Clarke1994-uj}, and the 2000s by a shift from running single programs to graphs of dependencies \cite{Deelman2019-il}. Two decades later, the 2020s are becoming defined by \gls{aiml} orchestration empowered by \gls{llm}s. It might be pointed out that the way in which we interact with machines is becoming closer to how we interact with one another. It is flexible, and conversational. It is not clear if this is advantageous due to expressibility, or deleterious due to loss of control. 

While a detailed analysis of the cultural and social implications are outside of the scope of this paper, we might summarize some of the broad challenges. The first challenge is representation of work, tasks, and outcomes. Traditional computing uses precise machine instructions \cite{ELF1997} in programs paired with return codes. The user goal is directly programmed, affording a tight level of control over program executions. The gap between the user goal and actions required to execute it grows with workflow tools that add a layer of abstraction to describe interactions and dependencies between tasks \cite{Deelman2019-il}. Adding \gls{llm}a moves the initial task to represent intent one step earlier. Instead of modeling our intent directly into software, we provide \gls{llm} agents with a goal that they must process to derive the entire execution, including inputs, tasks, and interaction with the environment. In this context, an ability to understand the environment and discover resources \cite{sochat2026agentic_science}  becomes paramount, along with setting and enforcing strong constraints on agent behavior. While the interactions are conversational, communication with an \gls{llm} agent is different than that with a human \cite{Lin2024-qi}. It cannot be assumed that agents will always make the correct or same decisions.

The next challenge is heterogeneity of resources, and complexity of scientific workflows. Whether it is a human or an agent executing tasks toward a goal, the higher degree of complexity of the resources leads to a greater number of possible configurations, which makes the task harder, and sometimes computationally infeasible \cite{Sharma2025-vn}. The inherent non-deterministic nature of agents makes their outputs often not reproducible \cite{Unknown2026-dq}. A lack of understanding that is created by way of allocating responsibility of ownership to \gls{llm}s is likely a barrier to establishing trust. Finally, there is a conflict between precision and flexibility. \gls{hpc} environments are unforgiving in the formats and commands they expect. Agents driven by \gls{llm}s are asked to achieve goals that are non-specific, such as optimizing a \gls{hpc} application. As an example, a prompt to schedule a workload might ask for optimizing for memory efficiency, or ensuring there is CPU affinity. An agent not only needs to understand the meaning of the requests, but to map the meaning to a specific workload manager that can provide it, and done exactly with the flags and arguments that the manager expects. Understanding the degree to which \gls{llm} agents can perform this task, and strategies for going about it, are the goals of this paper.

% @INPROCEEDINGS{Cummins2025-vb,
%  title     = "{LLM} compiler: Foundation language models for compiler
%               optimization",

In this work, we explore using descriptive language as more rigorous input for HPC optimization and agent-led job conversion. In our Method (Section \ref{sec:methods}) we demonstrate the ability of an agentic framework to design and run a scaling study across 5 well known \gls{hpc} applications and benchmarks, perform translation of job specifications between workload managers for Slurm and Flux Framework \cite{ahn-2014}, and design and execute bioscience workflows. We review results (Section \ref{sec:results}), demonstrating that an agentic framework can optimize and run \gls{hpc} scaling studies, translate the majority of job specifications between Slurm and Flux, and design and execute bioscience workflows that reproduce expert-written results with high fidelity, while revealing concrete, task-specific limitations in job monitoring, affinity translation, and adherence to specification. We make the following contributions:

\begin{compactitem}
 \item{Execution and analysis of 4 HPC applications in Kubernetes}
 \item{Best practices for \gls{hpc} translation prompts}
 \item{Agentic design and run of Snakemake workflows}
 \item{Evaluation of outcomes against human expertise}
 \item{Software prototype for agentic orchestration}
\end{compactitem}

We start with an introduction to agentic roles and definitions (Section \ref{sec:methods}) and describe our methods to orchestrate agentic execution and respond to failure. We describe experiments and results (Section \ref{sec:results}). We finish with a discussion of lessons learned and best practices for collaboration between agents and humans.
\section{Methods}
\label{sec:methods}

% TODO methods
% write up methods for scaling study
% write up methods for transform, talk about agents, need to summarize dataset (size?) and the breakdown of args used so we can say some don't work but it does not matter.
% limitation - what we left out.
% how we did in stages.

\label{sec:overview}
\subsection{Overview}

Our study includes four experiments that focus on real-world use cases. The first (Section \ref{sec:scaling-study}) is intended to test the ability of an agent to translate a human request and intent to run and scale an application into an orchestrated scaling study. The second experiment looked at job translation, testing an agent's ability to translate 1,000 job specifications from open source repositories into the Flux Framework workload manager (Section \ref{sec:translation}). The third experiment uses applications from our optimization study and our transformer agent. We vary performance flags across 1-5 nodes, and ask the translation agent to do an equivalent transformation to Slurm (\ref{sec:slurm}). We then test the scripts and assess performance and the agent's ability to predict negative implications of the translation.  Finally, we demonstrate design and execution of 3 Snakemake workflows, where the agent is provided with input data and a goal, and required to dynamically write and execute the workflow (Section \ref{sec:workflow-orchestration}).

\smallskip
\noindent{\bf Agentic Framework} 
\label{sec:agents}
An agentic framework is a design philosophy about how to combine interactions with an \gls{llm} with an initial user prompt, and tools. Specifically, the tools allow for exploration of the environment and discovery of resources, along with enabling the agent to execute interactions needed toward a task. We designed software, fractale \cite{vanessa_sochat_2026_19654652} that provides an agent expert abstraction \cite{vanessa_sochat_2026_19654729}, where the agent has expertise toward a scoped task. The expertise is defined by the prompt, and the agent is designed to work in a discovery, action, evaluation, and analysis loop, discovering tools available in the context of the task requested, and making turns of calls and analysis of the result until a desired outcome or a maximum number of attempts is reached. The software has a modular design for agents, allowing us to design experts for an optimization study, a job transformer, and a workflow orchestrator. The software supports agent backends for OpenAI family and Google Gemini models, and we chose to use Google Gemini for this work.

\subsection{Scaling Study} 
\label{sec:scaling-study}
We aimed to deploy 4 \gls{hpc} applications and benchmarks in the leading cloud orchestrator, Kubernetes, using previously built containers from prior work \cite{sochat2025agentic}. This task is motivated by our own experiences running scaling studies, for which the manual work is often arduous and time consuming, even with declarative management and automation \cite{sochat2025usabilityevaluationcloudhpc,Sochat2025-fp}. We chose Flux Framework \cite{ahn-2014} as our workload manager because it is easy to deploy in Kubernetes with a single manifest with the Flux Operator \cite{Sochat2024-the-flux-operator}, and our team has expertise in using it. Our applications include a set that are intended to strong scale (LAMMPS, Kripke), weak scale (AMG2023) or be used as benchmarks (OSU Benchmarks). We used the Amazon Web Services \emph{hpg7g.16xlarge} instance type with the \gls{efa} to enable low-latency networking \cite{amazon-efa}. For each application, we choose problem sizes that will run at approximately 5 minutes for the smallest number of nodes (N=1). We run 10 iterations for each size from 1 to 5 nodes, and have described each application in detail in previous work \cite{sochat2025usabilityevaluationcloudhpc}. 

\smallskip
\noindent{\bf Optimization Agent} 
\label{sec:optimization-agent}
The Flux Operator optimization agent is an expert at creation and deployment of the \emph{MiniCluster} \gls{crd}. To work successfully, it must authenticate with the cluster, discover resources available, generate and apply the \gls{crd}, and monitor the status of the job. An optimization agent goal should include a explicit task, constraints, criteria for completion, and other points of expertise from the human executioner. The agent must generate a command for Flux, along with ensuring the deployment matches the node.  The optimization agent can use any set of tools discovered via the \gls{mcp}. For our experiments, we deploy the expert alongside our mcp-server framework \cite{sochat2026agentic_science} that provisions 16 tools to interact with Kubernetes \cite{vanessa_sochat_2026_19654729} along with a Kubernetes event subscription.  

\smallskip
\noindent{\bf Agent Instructions} 
\label{sec:optimization-agent-instructions}
The optimization agent is given a 5 node cluster as resources. For each scaling study, we instruct the agent to discover cluster resources, and deploy a \emph{MiniCluster} \gls{crd} starting at size 1 that uses the maximum cores available. We instruct the agent to maximize or minimize a \gls{fom} relevant to each application at one node.  As an example, for LAMMPS, we instruct the agent to start at problem size x,y,z of 20 20 20, and then go up in increments of 1, and adjust individual sizes when \emph{OOMKilled}. When the run is successful and deemed optimized by the agent, we instruct the agent to pin problem size parameters and scale up to the maximum size of the cluster, 5 nodes, and report a final structured result in JSON. In practice, this means the agent must write and create the custom resource \emph{MiniCluster}, check status and debug errors that arise, clean up on error, wait for success, retrieve and parse logs to extract the \gls{fom}. The exception is with Kripke, for which it is not possible to deploy on 3 or 5 nodes, and we provide the agent with a 6 node cluster. We ask the agents to return \gls{fom}s, reasons for decisions, and literature references to justify them. For each of 5 applications and 5 sizes, we performed 5 scaling studies for a total of 25 studies and 125 size, application, and iteration combinations.

\subsection{Job Translation} 
\label{sec:translation}
A second challenge for \gls{hpc} practitioners is translation of jobs to run between workload managers. Moving work between managers has become an increasingly common task given resource contention and the need to use multi-cluster environments. For this experiment, we start with an open source database of 33,744 job specifications parsed from GitHub \cite{vanessa_sochat_2026_20112396} and remove outliers outside of the 95th percentile ($N=32,001$). We randomly select 1,000 job specifications with 1,000 or fewer tokens to be more conservative with respect to sending tokens to the Gemini API. To ensure the sample is comparable in complexity to the larger set, we calculate the cyclomatic complexity using \emph{shellmetrics} \cite{1702388}. The cyclomatic complexity is a quantitative measure of the number of linearly independent paths through a program's source code. Importantly, our 1K subset ($N=1,000$, $\mu = 1.16$, $\sigma = 0.70$) is comparable to the larger sample ($N=32,001$, $\mu = 1.64$, $\sigma = 1.31$), with a Z-score of $-0.369$ (within $0.37$ standard deviations of the larger sample mean).

% We deploy the expert alongside our mcp-server framework \cite{sochat2026agentic_science} that provisions 16 tools to interact with Kubernetes \cite{vanessa_sochat_2026_19654729} along with a Kubernetes event subscription.  

\smallskip
\noindent{\bf Job Translation Instructions} 
\label{sec:translation-instructions}
For each transformation, we provide the agent with full help output for the Flux submit command, and the instruction for how to represent a flag in a batch script. % We previously tested an approach to instead provide the agent with a tool to call a Flux validator, but continually found bugs in the software that deemed it not reliable to use for these experiments.
We additionally tested a more detailed case, asking the agent to ignore accounts, partitions, queues, mail, banks, notifications, and commented out directives. We also provide clear instruction for how Flux models affinity after observing mixing up with Slurm affinity options.

% Note to dan - this is an entire other area of work so I am commenting / leaving it out. Too much already.
\begin{comment}
\smallskip
\noindent{\bf Result Discovery} 
\label{sec:result-discovery}
Our next experiment is driven by the need to parse a large set of files and find relevant information or data. As an example, our results from the scaling study are organized into timestamped directories with named files for events and metrics. To find a final result requires finding the right file, and parsing the result for the last entry in a structured JSON. Arguably, agents can help with this task to derive a final result. For this experiment, we created a content discovery expert in our fractale agents library. The discovery agent is provided with a goal, and given read access to a root with data. The agent is asked to:

1. Discover and summarize the structure of results and applications found
2. For each application, look at the file types, and write a script that will read in and summarize sections found in the file.
3. Using this script, look for the final report of the FOM at each size.
4. Once you discover the structure, apply to all files and generate one summary (each result, etc.)
\end{comment}

\smallskip
\noindent{\bf Job Translation Agent} 
\label{sec:translation-agent}
The job translation agent is an expert at transforming job specifications between workload managers, honoring a user request for a source and destination manager, along with user-specific goals. The agent is instructed to preserve as many options as possible from the original, and leave out directives have no means to translate, with comment about performance implications. For our translation task, we request to leave out directives related to user or cluster identity such as queue or partition and email notifications. We instruct the agent that if a validation tool is available, it should be used to validate the job specification. Finally, the agent is instructed to derive a list of issues with labels \emph{UNKNOWN\_TO\_ME} (a parameter not familiar with that cannot be figured out), \emph{NO\_ANALOGOUS}, and \emph{MISSING} for when there is no possible conversion.

% Note that the Flux validator is not used here because we are validating slurm. I also am finding more bugs with it 

\smallskip
\subsection{Translation Performance} 
\label{sec:slurm}
To combine our first two experiments, we use the final optimized container images from our scaling and optimization study (Section \ref{sec:scaling-study}) and ask the transformer agent to transform each of Kripke, LAMMPS, AMG, and OSU Benchmarks \emph{osu\_latency} and \emph{osu\_allreduce} from Flux into Slurm. For each, we generated and requested job specifications between 1 and 5 nodes, with and without affinity. We ran 10 iterations for each application at each size, and for with and without affinity. Our next goal was to test the actual performance of the translated job specifications compared to the original. For our original job specifications, we again used the Flux Operator \cite{Sochat2024-the-flux-operator}, this time manually orchestrating the work from 1 to 5 nodes.  For Slurm, although we first intended to use AWS Parallel Cluster or AWS Parallel Compute Service with Singularity \cite{kurtzer2017singularity} containers, we were unable to deploy the setups customized for our needs. As an alternative, and a solution closer to the Flux Operator in using a deployment of the workload manager directly in Kubernetes with the same containers, we used the Slurm Operator for Kubernetes \cite{vanessa_sochat_2026_20128848}. A successful translation result will be indicated if the performance between runs at equivalent scales and affinity are not significantly different.

\subsection{Workflow Orchestration} 
\label{sec:workflow-orchestration}
A final challenge for descriptive execution is design and orchestration of workflows.  The Snakemake workflow manager \cite{Molder2021-rq} is ideal for automation by an agent because it defines Snakemake wrappers -- packaged single unit steps that are common across bioinformatics workflows. By way of designing a catalog of Snakemake tools for the agent via our server and providers software \cite{vanessa_sochat_2026_19476501}, the agent is able to design the and execute the entire workflow. The mcp-server is deployed in the context of a job, with a specific scope to execute one workflow. A read only data input directory is provided alongside a read and write working directory. When the server starts, it stages input data and workflow steps to the working directory for the agent to work from. It dynamically downloads and parses the snakemake wrappers repository. Traditionally, a human might write the workflow file that defines steps with inputs and outputs, called a Snakefile. In this case, the agent will be tasked to write it, and provided with functions to view the current Snakefile, roll back a step, or get metadata about a specific wrapper. Upon startup, the agent receives a prompt with a specific goal (e.g., variant calling or sequence alignment) and is able to see a suite of tools to interact with snakemake, including listing and getting details for wrappers, viewing or editing the current Snakefile, inspecting data and inputs, and executing wrappers or rules. The staging design is ideal as it protects the agent from destructing actions against input data, and allows for multiple agents to work on the data at once. There is no redundancy because the data is linked, and the links can be deleted without harming the initial data. The software rolls back failed steps automatically, which includes removing the Snakefile rule and the step output directory. To assess the ability of an agent to orchestrate a workflow, we run a Snakemake variant calling workflow 20 times and compare against a ground truth, manually generated Snakefile.

\smallskip
\noindent{\bf Snakemake Agent} 
\label{sec:snakemake-agent}
The Snakemake workflow expert is given input data, and tasked with   a scientific objective. It must execute a complete workflow until the goal is reached. The agent is given a structured input directory, permissions and path details, and instructed to inspect the computational environment to inspect wrappers and data. For execution, the agent adds and executes one step at a time and ensures success or makes another attempt. A structured JSON object must be returned upon completion. During the prototyping stage, We tested us rolling back and agent rolling back, and allowing agent to explicitly delete step. The latter worked better and we roll back failed steps. Rollbacks are immutable. 

\smallskip
\noindent{\bf Snakemake Analysis} 
\label{sec:snakemake-analysis}
Our analysis goal is to compare agent-generated workflow results against a manually written
reference pipeline that can be considered a ground truth. The reference pipeline maps short reads to a reference genome, sorts and indexes the alignments, and jointly calls genomic variants across three samples.
We are careful to install the same \textit{samtools} and \textit{bcftools} as in the wrappers to ensure comparability of output -- any deviance in results can be attributable to the workflow orchestration. For each output iteration, the agent will generate a full output directory and \emph{Snakefile}. Because the agent decides the specific steps to use and the naming of the output step directories, there can be variation in the output structure, and so the analysis does not assume a fixed directory layout or file paths.

For each run we locate the final variant-call file based on content rather than path. We classify each candidate file using the provenance \texttt{bcftools} writes into its output headers 
% (e.g. \texttt{bcftools\_callCommand}, \texttt{bcftoolsCommand=mpileup}) 
and the presence of a genotype (\texttt{GT}) field. A file produced by \texttt{bcftools call} is taken as a call set, whereas a file produced only by \texttt{bcftools mpileup}, lacking genotypes, is treated as an intermediate. When multiple call sets are present, the one consumed as input by another based on input paths embedded in the header is treated as an upstream. Each call set is left aligned and decomposed into biallelic records against the reference (\texttt{bcftools norm -f ref -m -any}), sorted, and restricted to variant sites (records with an alternate allele), so that the reference output (\texttt{bcftools call -mv}, variant sites only) and the agentic output (\texttt{bcftools call -m}, all covered sites) are directly comparable. We intersect the reference and agentic call sets with \texttt{bcftools isec}, partitioning sites into shared (true positives, TP), reference-only (false negatives, FN), and agentic-only (false positives, FP), with the reference taken as truth. From these we compute precision, recall, $F_1$, and the Jaccard index. A run is an exact reproduction when $\mathrm{FP}=\mathrm{FN}=0$. Comparison is at the site level (\textsc{chrom}/\textsc{pos}/\textsc{ref}/\textsc{alt}) and is therefore independent of sample labels. Joint calling is performed across the sample set defined by the reference. The agent's sample set is recorded for each run, and where a run carries a different set, its call set is restricted to the samples common with the reference (\texttt{bcftools view -s\,$\langle$samples$\rangle$ --trim-alt-alleles}) before comparison. For runs that apply a step after calling (e.g.\ a hard-quality filter) absent from the reference, we report the concordance of the terminal deliverable and, separately, of the upstream call-stage output.
 
We may find that we arrive at identical results haven taken different paths, or using a different structure of steps. We would want to understand the overall categorization of steps. To characterize how each workflow is constructed, we parse the per-run step directories and rule definitions and mapped each step to a canonical scientific operation (alignment, sorting, indexing, pileup, calling, and optional steps such as
filtering, normalization, merging, and read-group assignment) by keyword. For each run we record the number of steps, the sequence of canonical operations, whether rules were expressed via Snakemake wrappers or shell commands, and the presence of any additional operations. Across runs we count the distinct step-name layouts, the distinct canonical operation sequences, and, per operation, the number of distinct step-directory names used to implement it.  The analysis is implemented as command-line tools depending only on \texttt{bcftools} and a \texttt{faidx}-indexed reference.
\section{Results}
\label{sec:results}

% scaling study plots and evaluate literature references
% write up results (results and images)
% look at workflow output, figure out what we can tweak
% choose more workflows, run, parse results
% need to figure out graph analyses to include

\subsection{Scaling Study}
\label{sec:results-scaling-study}

\begin{figure*}[ht]
    \includesvg[width=1.0\linewidth]{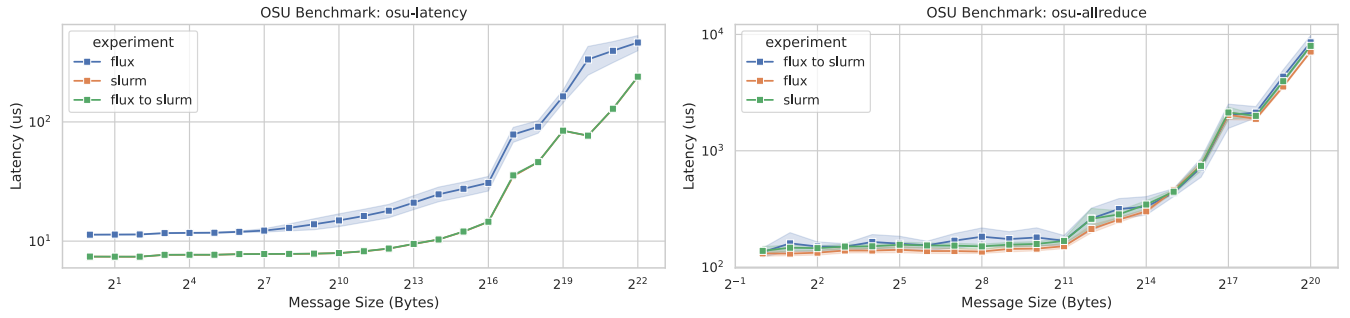}    
    \centering
    \vspace{1em}
  \caption{Translation performance for OSU Benchmarks \normalfont. OSU Latency results between Slurm and Flux to Slurm were almost identical, and overlap on the plot. OSU All Reduce is reported across message sizes for 5 nodes.}
  \label{fig:osu-performance}
\end{figure*}

\begin{figure*}[ht]
  \centering
  \includesvg[width=2.0\columnwidth]{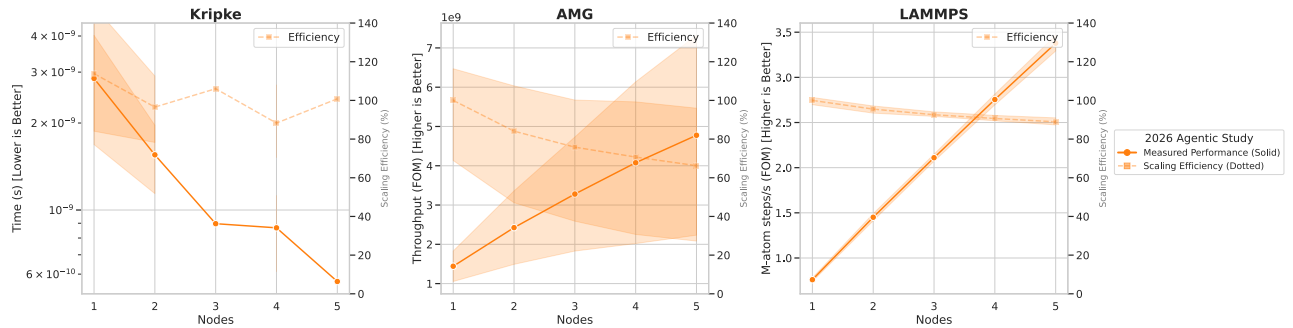}
  \caption{\normalfont Maximum \gls{fom} achieved for HPC applications for 1-5 nodes.}
  \label{fig:scaling-study}
\end{figure*}

We performed a scaling study across five nodes and three applications, asking the agent to optimize at the smallest size and scale up appropriately to maximize or minimize a \gls{fom} of interest. Best achieved \gls{fom} for each optimization size are reported in Table \ref{tab:performance_results} and results across independent studies shown in Figure \ref{fig:scaling-study}. For LAMMPS, variance was exceptionally small, reflecting consistency in the runs. Kripke was the most erroneous, with the majority of the runs being unsuccessful to deploy the application to an odd number of nodes. For all applications, we were predominantly interested in the agent's ability to perform the study and adjust parameters when required. The variance for AMG is large due to independent scaling studies arriving at different maximum problem sizes. An interface to explore full results is available \footnote{https://converged-computing.org/fractale-experiments/optimization-study/data/single-node-1/}.

% NOTE TO DAN: As a person / human I am having a hard time in the interface to cleanly summarize the results. A lot of the variation is due to different problem sizes. Should we be providing the problem size? Should we plot with some kind of normalization? The plot seems misleading / wrong due to this. Of our 10 runs, not all were perfect. E.g., sometimes it would fail to parse a log. Or deploy a size. We need to report on that. We also need to better capture the agents explicit choices for each app problem size and params, so we have something meaningful to say. I am overall not happy with this experiment result, and I think we need to run this study again, as much as I hate to say it. But with discussion about how to do it, better controlling / modeling what we want to capture. When I ran these I largely did not have any specific goals.  I want you to take a look first.

% We found that the expert agent could better perform during optimization when instructed to  start at smaller problem sizes and scale up.  % If an application too quickly exhausts memory (\emph{OOMKilled}) it could often attribute the error to something else. or restarts that were not caught, leading to container statuses of unknown. The agent instructions

\begin{table}[ht]
    \centering
    \caption{Best Achieved \gls{fom} Performance}
    \label{tab:performance_results}
    \begin{tabular}{llcc}
        \toprule
        \textbf{Application} & \textbf{Metric} & \textbf{Best Value} & \textbf{Nodes} \\
        \midrule
        Kripke & Time (s) & $5.6584 \times 10^{-10}$ & 5 \\
        AMG    & \gls{fom} Overall & $6.9176 \times 10^{09}$  & 5 \\
        LAMMPS & M-atom steps/s & $3.5010$  & 5 \\
        \bottomrule
    \end{tabular}
\end{table}

% Successful Run Counts per Configuration (2026)
% =================================================================
% Nodes   1  2  3  4  5
% App                  
% AMG     3  3  3  3  3
% Kripke  5  5  1  5  1
% LAMMPS  6  6  5  5  4

% Node 3: Successful run found (Time: 8.9677e-10). Claimed constraint is FALSE.
% Node 5: Successful run found (Time: 5.6584e-10). Claimed constraint is FALSE.

% FINAL REPORT: Best Achieved Performance (2026)
% Application     | Metric       | Best Value   | Node
% Kripke          | Time (s)     |   5.6584e-10 | 5
% AMG             | FOM          |   6.9176e+09 | 5
% LAMMPS          | FOM          |   3.5010e+00 | 5

\subsection{Job Translation}
\label{sec:results-translation}

\begin{figure}[hb]
  \centering
  \includesvg[width=1.0\columnwidth]{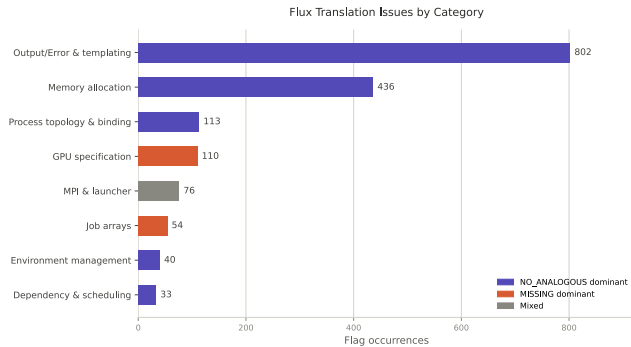}
  \caption{\normalfont Categorized issues for Flux conversions. Flux does not have a memory flag, and the agent }
  \label{fig:translation-issues}
\end{figure}

Of the sample, 18 jobs were not converted due to not being recognizable batch scripts associated with a workload manager, leaving 982 jobs. Of the 982 jobs, all translations were validated by production Flux, meaning successful submission of a dry run without report of error. Importantly, a validation does not indicate that the translation was able to reliably map the intention of the job specification. Across translations, the agent reported no analogous translation 1033 times, s missing parameter 237 times, and an unknown directive 62 times. We generated 8 categories of issues based on groups of flags reported problematic by the agent, including I/O, memory, topology, GPU specification, MPI, job arrays, environment, and scheduling Figure \ref{fig:translation-issues}. The translation agent has the most difficulty with I/O (N=802 issues), primarily due to being unable to handle Slurm templates for input and output files translated to Flux. While important for preserving data in real world runs, these flags with no direct analogous mapping do not influence application performance.  Memory allocation (N=436) was an expected result, as Flux deploys predominantly to clusters that require exclusive node access (all memory) and has no flags to specify memory for a job. Flags related to slurm \emph{gres} were noted as problematic issues by agents (N=110), along with a lack of job arrays (N=54). An interface with complete results for exploration is available \footnote{https://converged-computing.org/fractale-experiments/jobspec-agent-conversion/results/}.

% Closer inspection of this data revealed that the biggest issue with translation to Flux was missing flags for memory \emph{--mem} (N=355 of the no analogous, or 34.4\%) and \emph{--mem-per-cpu} (N=183 or 17.7\%). The agent also marked erroneous flags that indicate any kind of templating (e.g., \emph{\%j} or \emph{--output} that uses it () for output names

% Note from V_ I am adding up %j, --output, -o

% TODO need to look at this and synthesize as a person.

% 1377 of 1378
% ===========================================================% =
% Experiment                | Success  | Failure 
% ------------------------------------------------------------
% 1k                        | 982      | 0       
% base                      | 198      | 0       
% details                   | 198      | 0       
% ============================================================

% https://converged-computing.org/fractale-experiments/jobspec-agent-conversion/results/

\subsection{Translation Performance}
\label{sec:results-translation-performance}

\begin{figure*}[ht]
    \includesvg[width=1.0\linewidth]{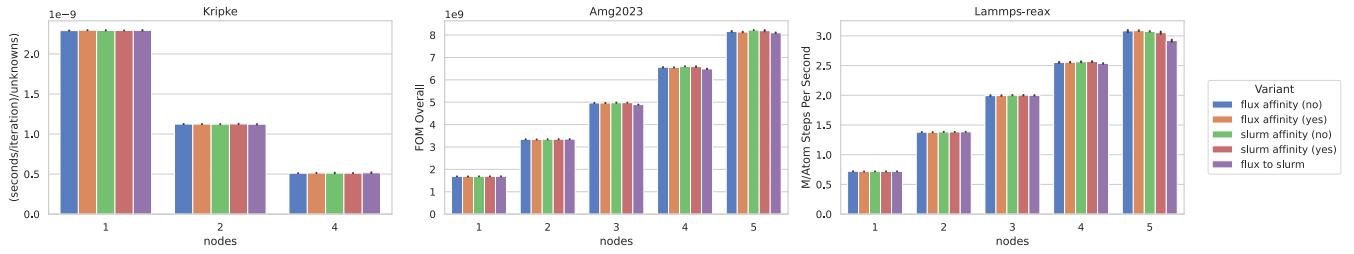}    
    \centering
    \vspace{1em}
  \caption{Translation performance \normalfont for a scaling study across up to 5 sizes for two affinity variants for each of Flux and Slurm. There are no observable differences with respect to adding affinity. Asking the agent to}
  \label{fig:translation-performance}
\end{figure*}

We used application containers from our scaling study with each of Slurm and Flux operators in Kubernetes to run job commands generated by human experts as compared to an agentic translation. Specifically, we did a conversion from Flux to Slurm. Results for each of AMG, LAMMPS, and Kripke are reported in Figure \ref{fig:translation-performance} and for OSU Benchmarks in \ref{fig:osu-performance}.  The performance of the agent translation from Flux to Slurm at 5 nodes for  \emph{AllReduce} ($\mu=11645.25, \sigma=4237.68$) showed a 28.79\% increase in mean latency compared to human-derived Slurm ($\mu=9042.21, \sigma=353.3$). Due to high performance variability, the difference was not statistically significant ($p=3.99e-01$). There were no significant differences for any other application between Flux, Slurm, and the Flux to Slurm translation, with the exception at LAMMPS at size 5 for the Flux to Slurm translation ($\mu=2.920, \sigma=0.03$) as compared to Slurm with and without affinity ($\mu=3.06, \sigma=0.03$). For the set of translations at this size, jobs that attempted to set affinity were not successful, meaning that the result only includes jobs without affinity. All other sizes have at least one affinity run.  Lack of affinity would likely be more apparent at larger sizes.

%In [31]: subset.groupby(['experiment']).fom.mean()
%experiment
%flux             3.085250
%flux to slurm    2.920222
%slurm            3.064125
%Name: fom, dtype: float64

%In [32]: subset.groupby(['experiment']).fom.std()
%Out[32]: 
%experiment
%flux             0.026463
%flux to slurm    0.028687
%slurm            0.026518
%Name: fom, dtype: float64

% size 5 nodes for lammps
% improve-4 uses --distribution=cyclic (round-robin rank scatter, no explicit tpn) plus --exclusive. improve-9 uses --ntasks-per-node=64 and --cpu-bind=cores but no distribution or exclusive. At 4N where improve-8 combines all of them: block:block + exclusive + cpu-bind=cores + tpn=64. 5N improved commands are individually less complete than at other scales.

Closer inspection of all applications and benchmarks revealed performance differences were due to not asking Slurm correctly to set affinity. For a total of 76 Slurm commands across 5 applications between 64 and 320 tasks, 63 ran successfully, and 13 produced erroneous \emph{cpu-bind} flags, asking for one of \emph{task}, \emph{tasks}, or the Flux flag \emph{per-task}. Looking at agent translation logs, in all erroneous cases the agent identified the Flux flag as not having an analogous translation, but added the incorrect flag anyway. This addition violated an instruction given to our agent. We believe this results from the agent getting caught up on the discrepancy between \emph{cpu-affinity} being an option flag, and Slurm not having option flags. In assessing this discrepancy, the agent forgets the initial prompt to not hallucinate flags or options.
Finally, the most interesting result was the discrepancy in performance between Slurm ($mu=7.41,\sigma=0.03$) and Flux ($mu=11.32,\sigma=0.22$) for \emph{osu-latency} shown as the gap between blue and green in Figure \ref{fig:osu-performance}.

\subsection{Workflow Orchestration}
\label{sec:results-workflow-orchestration}

We ran a full workflow for variant calling with Snakemake using a Snakemake agent and compared to 21 iterations of an equivalent ground truth run. The agent followed a similar pattern of behavior each time. We observe the agent looking at the environment, specifically the data input directory, wrappers available, and metadata about wrappers of interest. We then see the agent calling \emph{bcftools} for variant calling for each of three samples. The agent is able to see the Snakefile on request, delete or execute rules or wrappers, and list contents of the input or working directory. 

\begin{figure}[bh!]
  \centering
  \includegraphics[width=1.0\linewidth]{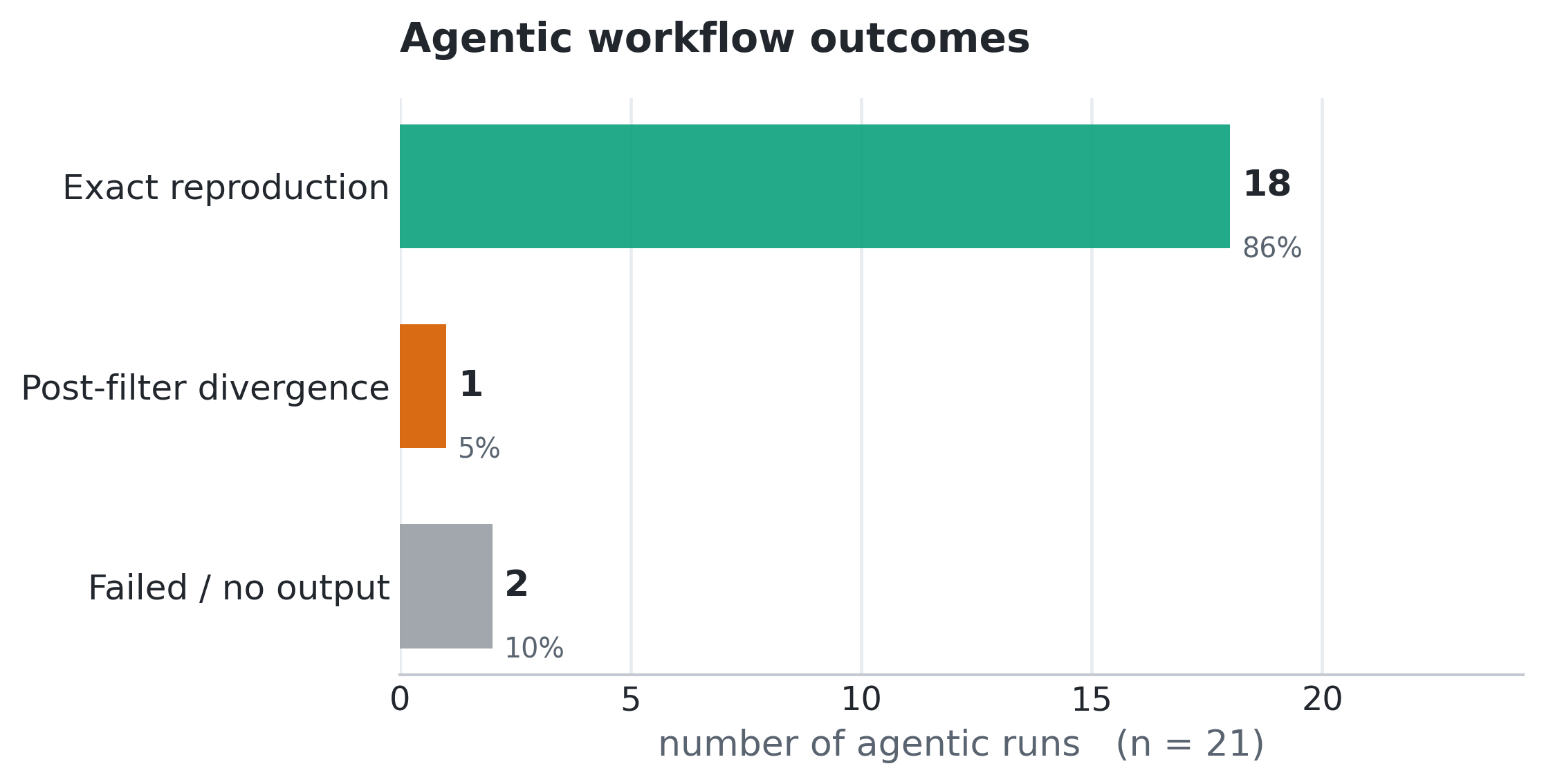}
  \caption{Outcomes across 21 agentic runs. \normalfont Exact reproduction of the reference call set, post-filter divergence, and runs producing no output. Percentages are of all runs.}
  \label{fig:outcomes}
\end{figure}

\begin{figure}[b]
  \centering
\includegraphics[width=1.0\linewidth]{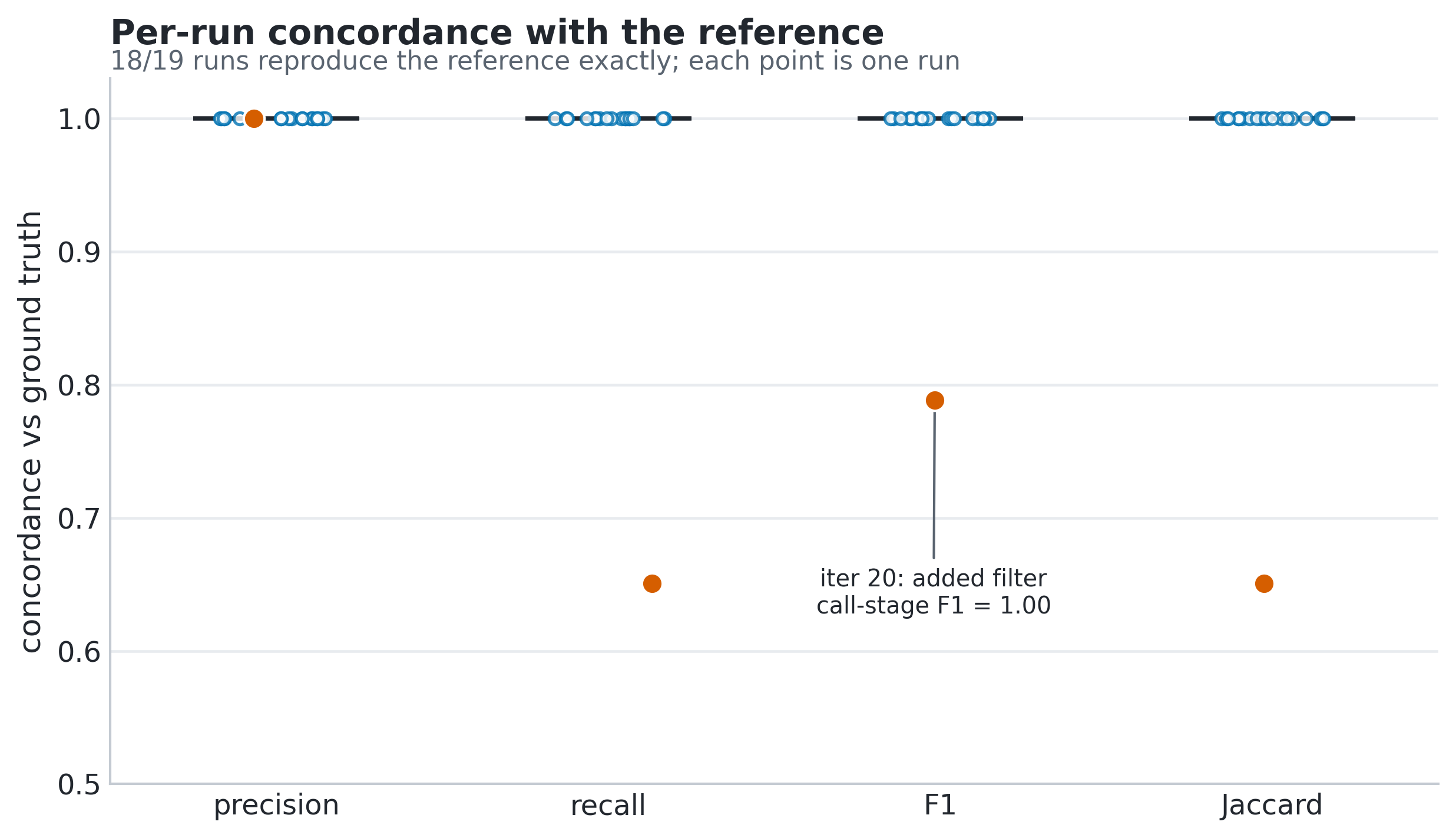}
  \caption{Per-run concordance with the reference. \normalfont We show precision, recall, $F_1$, and the Jaccard index.  Each point is one run and boxes summarise the distribution. Almost all runs sit at $1.0$. The single outlier (iteration~20) added a post-calling filter and is otherwise concordant at the calling stage.}
  \label{fig:concordance}
\end{figure}

Of $21$ agentic iterations, $19$ produced a usable variant call set. The remaining two did not (one resulted in a partial workflow with no final call set likely due to connectivity error, and the other was not successful to produce output). These runs were excluded from concordance analysis (Figure ~\ref{fig:outcomes}). Of the $19$ runs we could evaluate, $18$ ($95\%$) reproduced the reference call set exactly (zero false positives and zero false negatives), giving a mean $F_1$ of $0.989$ and mean Jaccard of $0.982$ against the reference. Per-run concordance was at the ceiling for precision, recall, $F_1$, and the Jaccard index, with a single exception (Figure ~\ref{fig:concordance}).

The lone divergent run (iteration~$20$) applied a hard-quality filter after calling that the reference workflow did not. This reduced its recall (terminal $F_1=0.79$) by removing $230$ true variant sites, while its precision remained $1.0$. Its upstream call-stage output was perfectly concordant with the reference (call-stage $F_1=1.00$), localizing the discrepancy to the added filter rather than to the variant caller. Because all runs that we were able to evaluate processed the reference sample set, the as-produced and sample-matched comparisons were identical, and no run was penalized for sample-set differences.

Although the scientific result was almost invariant, the workflow \emph{implementations} were not. Every run performed the core operation graph: alignment was present in $100\%$ of the $20$ runs that composed a workflow, and sorting, indexing, pileup, and calling each in $90$--$95\%$ (Table \ref{tab:ops}). The small shortfalls reflect the partial run that produced no call set together with occasional folding of an operation into an adjacent step (e.g. sorting within another rule, or fusion of pileup and calling). Only a single run introduced an additional operation (a post-calling filter, and one run a merge step; $1/20$ each). At the same time, the surface form varied substantially  runs used between $3$ and $11$ steps (median $5$), and a given operation was expressed under up to $6$ distinct step names across runs (Table \ref{tab:ops}). In aggregate the $20$ runs exhibited $10$ distinct step-name layouts but only $7$ distinct canonical operation sequences (Table \ref{tab:structure}), indicating that the agent's variability is concentrated in naming and organization rather than in the scientific procedure.

\begin{table}[b!]
  \centering
  \caption{Canonical scientific operations across agentic runs that produced a workflow. \normalfont \emph{Runs} counts those performing the operation, and \emph{Names} is the number of distinct step-directory names used to implement it.}
  \label{tab:ops}
  \begin{tabular}{lcc}
    \toprule
    Operation & Runs (of 20) & Distinct names \\
    \midrule
    \multicolumn{3}{l}{\textit{Core operations}}\\
    Alignment & 20 (100\%) & 6 \\
    Sorting & 18 (90\%) & 5 \\
    Indexing & 19 (95\%) & 6 \\
    Pileup & 19 (95\%) & 3 \\
    Calling & 19 (95\%) & 3 \\
    \midrule
    \multicolumn{3}{l}{\textit{Additional operations}}\\
    Filtering & 1 (5\%) & 1 \\
    Merging & 1 (5\%) & 1 \\
    \bottomrule
  \end{tabular}
\end{table}

Taken together, the agent reproduces the reference scientific result with high reliability (exact reproduction in $95\%$ of runs we could evaluate), and its few deviations are workflow-design choices---an added filtering step in one run, and failure to produce a usable call set in two---rather than errors in the underlying analysis. 

\begin{table}[t]
  \centering
  \caption{Structural variation across the 20 agentic workflows.}
  \label{tab:structure}
  \begin{tabular}{lr}
    \toprule
    Quantity & Value \\
    \midrule
    Runs producing a workflow & 20 \\
    Steps per workflow (min / median / max) & 3 / 5 / 11 \\
    Distinct step-name layouts & 10 \\
    Distinct operation sequences & 7 \\
    Distinct operation sets & 5 \\
    Runs using wrapper / shell rules & 20 / 4 \\
    Runs adding a filter step & 1 \\
    \bottomrule
  \end{tabular}
\end{table}

\section{Discussion}
\label{sec:discussion}

% agent typically chose event subscription, 

In this work, we use expert agents to semi-autonomously deploy and optimize \gls{hpc} applications, convert job specifications between managers, and orchestrate workflows. We use the same underlying orchestration software with different expert agents across tasks. There are several topics for discussion.

\subsection{Scaling Study}

The variability of a \gls{fom} (AMG) or inability to deploy an application at a particular node size (Kripke) reflects an inadequacy of tools to expose the environment and application deployment difficulty. A tool that might provide hints or validation about parameter choices would avoid deployment, wait, and failure of erroneous commands. The serial design of our scaling study reflects a need to monitor execution, and is not representative of an ideal execution environment to run jobs in parallel with a workload manager queue. For our experiments, an agent running in an asynchronous loop to execute, wait for, and retrieve logs for an application must know when a job is complete. The task of waiting, in retrospect, led to many errors by agents concluding an incomplete log indicated an erroneous job when it required more execution time.  While we provided and provisioned event notifications and gave the agent the ability to subscribe to job events, our expert agents more often chose to call an explicit sleep function between job executions, with increasing time as an experiment progressed. This setup does not afford optimal utilization of resources as agents often overestimate waiting time. A more intelligent design would use models to predict expected running times, and enforce the agents to subscribe to events. Subscribing to events would make it easer for one agent to handle more than application execution at once given a workload manager that delivers notifications across jobs. We also ran the study from a tabula rasa each time without giving agents additional hints about errors or preserving learning across iterations. An agent that is able to use a tool that exposes a model or database of previously successful runs could more quickly get an application working, and then better optimize.  Further, a more realistic scenario is one of agent managing multiple jobs, and needing to explicitly respond to events.  Providing tools and models that make it possible to better estimate job running times is not just important for checking status, but also for the initial scheduling. While we are actively working on parallel execution with events, we chose a serial design for this early study to make it easy and possible for the developer human user to observe the experiments. 

The serial task of performing optimization via submitting and waiting for jobs often does not reflect the human experience of testing applications interactively. To compare to what a human would do, experiments might test giving an agent access to the same \emph{MiniCluster} from the lead broker, akin to running the application on bare metal across an interactive allocation. Without needing to wait for wrapping orchestration of Kubernetes, for example, the optimization and subsequent experiments could be run much faster. 

% NOTE TO DAN: Let's discuss what would be ideal to run this. I don't know if a user is actually going to give an agent a random container and YOLO - go figure it out! But that is what we are testing here. We might want to do something more scoped and specific, but everything I can think of is better scoped to Heuretes. The best way to do this would extend to using events with our mcp server stuff, but that extends into hueretes and I don't want to bring it into here. So I think we should report and apologize for the serial orchestration and call it a day.

% What is missing here for me is being able to say "I want to run X IN THIS SPECIFIC WAY" and then having the agent figure out the right resources / jobspec. That is more of a dispatch experiment though, which is out of scope for this paper (we have another dispatch experiment I started writing up).

% For our scaling study we ran the iterations separately, meaning the model had no knowledge of the scaling study it just did. in a real world use case, we would be able to give the model information about previous attempts, and skip all the testing / failure of optimization at the begining. it would start in a b etter state. We did not do that here because we wanted to see if the blank tabula rasa would reproduce.

\subsection{Performance Translation}

The agent did well to translate jobs between workload managers Flux and Slurm, with the exception of affinity flags. Even with help provided for each of Slurm and Flux in the prompt, providing an incorrect flag option was common. The translation agent would often make more errors when distracted with an irrelevant detail. For affinity, Flux uses broker flags that are requested with an \emph{-o} option. In cases where the agent made a mistake, it commented verbosely on the inability to map the broker option to Slurm, and then subsequently hallucinated an incorrect affinity flag. We think that this issue might be resolved given a Slurm validator tool or an adversarial agent approach where one agent derives a command and a second is required to scrutinize it. A surprising and pleasant result in our work was observing relative equality of performance between orchestration of applications using Flux and Slurm. In a highly contentious landscape of workload managers, we can attest from our work that both are good choices in this set.

\subsection{Workflow Design and Reproduction}
The biosciences task asked the agent to design and execute a complete variant-calling workflow---read alignment, sorting, indexing, and joint variant calling-rather than to fill in a fixed template. Assessed against a manually-written reference, the agent was  reliable. and a single divergence attributable to an extra hard-filtering step the agent added of its own accord rather than to an error in calling.  What the agent did \emph{not} do reliably was converge on a single implementation, exhibiting ten distinct step layouts but only seven distinct operation sequences. The framework thus reproduces the scientific result more consistently than the \emph{code} that produces it.

This separation between functional and surface variability has a practical consequence for how agentic workflows should be evaluated. Comparing agent output to a reference by file paths, filenames, or even rule text is misleading, because two correct workflows rarely share them. We found it necessary to compare the scientific artifacts directly---normalizing and intersecting the called variant sites---and to recover each workflow's
structure from tool provenance rather than from naming.  While we expect this content- and provenance-based evaluation to generalize to other scientific workflows, we recognize that this task would be challenging given result files that cannot be compared.

Finally, the deviations we observed---an unrequested filtering step and two
empty outputs---are similar to failure we saw in other experiments in this larger study.  Akin to the hallucinated affinity flags in translation, an agent that adds or omits a step would benefit from a validation tool or an adversarial reviewer that checks the produced workflow against the specification before it is run.

\subsection{Limitations}

Our findings are conditioned on a single model and version, a fixed prompt design, and a particular tool set. Other LLM \gls{api} endpoints, newer model versions, or different prompt and tool configurations could shift both accuracy and failure modes, and we did not test across these choices. Agent behavior is also sensitive to prompt phrasing in ways we observed but did not quantify. For example, the translation agent made more errors when distracted by irrelevant detail. Absolute error rates should be read as
characteristic rather than definitive.

Each task was evaluated at experimental scale, including a serial scaling study, a bounded set of translation cases, and 21 independent bioscience workflow designs. These are sufficient to explore behavior and failure states, but are too few to report tight confidence intervals or to claim that rare failures have been observed. The serial, human-observable execution design further means our execution is not optimized. An optimized parallel, event-driven deployment would be ideal. As agentic approaches mature, we expect execution to be better optimized.

Our agentic outputs are assessed against references and a small set of well-known applications, benchmarks, and a tutorial-derived biosciences pipeline. A single expert reference is just one way to perform a task. An agent output that diverges may be wrong, or merely different and still correct. Our concordance metrics treat the reference as ground truth when in fact there may be multiple derivatives of \emph{correct}. Our experiment tasks themselves are representative but narrow, including five \gls{hpc} applications, two workload managers, and one variant-calling workflow.  Generalization to larger, multi-step, or less-documented scientific workflows remains to be shown. The biosciences comparison is also a site-level assessment of the final call set, and the workflow runs very quickly. We did not evaluate genotype-level concordance, runtime, or cost, which may matter for other workflows.

We ran each experiment from a blank slate, without preserving learning across iterations or supplying error hints, and with a human able to observe but largely not intervene. This strategy does not measure the more realistic collaborative setting in which an agent accumulates context, manages multiple concurrent jobs, and escalates to a human on uncertainty. We do not yet know when an agent should defer to a human, nor did we measure how much the failures we report (e.g. hallucinated affinity flags or an unrequested filtering step) would be caught by a validation or adversarial-review mechanism. Finally, allowing semi-autonomous agents to deploy and execute code at \gls{hpc} scale raises safety, cost, and reproducibility concerns. 

\section{Conclusion}
Across three tasks, our experiments show that agentic orchestration is already reliable enough to be useful and revealing enough to be improved. Agents optimized and ran scaling studies, translated the majority of job specifications between Slurm and Flux, and reproduced an expert-designed biosciences workflow while exploring a wide space of equivalent implementations. The errors that remain, including inefficient job monitoring, flag hallucination, and occasional departures from specification, are concrete and addressable through validation tools, event-driven execution, and persistent memory of past runs.

Agentic frameworks represent a transformative shift for the \gls{hpc} community and the broader software landscape. The realization of autonomous, converged \gls{hpc} infrastructure enables a collaborative paradigm where machines and humans jointly orchestrate the full lifecycle of scientific workloads. By integrating autonomous task execution and \gls{ai}-optimized scheduling, these advancements will fundamentally redefine the workflow of scientists and engineers. We are entering a new era of \gls{hpc} -- one defined by unprecedented speed, precision, and accessibility.

\begin{acks}
We are grateful for new cities and mountains to explore, colorful climbing walls, bows in hair, and the color green. This work was performed under the auspices of the U.S. Department of Energy by Lawrence Livermore National Laboratory under Contract DE-AC52-07NA27344 and was supported by the LLNL-LDRD Program under Projects No. 24-SI-005 (LLNL-CONF-2021324).
\end{acks}

\bibliographystyle{ACM-Reference-Format}
\bibliography{references}

% \appendix
% \input{sections/appendix}

\end{document}